\documentclass[aps,singlecolumn,showpacs,preprintnumbers, twocolumn]{revtex4-2}
\usepackage{float}
\usepackage{dblfloatfix}
\usepackage{fixltx2e}
\restylefloat{table}
\usepackage{graphicx}  % Include figure files
\usepackage{subfig}
\usepackage{subfloat}
\usepackage{multirow}
\usepackage{fancyhdr}
\usepackage{parskip}
\usepackage[T1]{fontenc}
\usepackage{dcolumn}   % Align table columns on decimal point
\usepackage{bm}        % bold math
\usepackage{amsfonts}  % Common math fonts
\usepackage{amsmath}   % Common math functions
\usepackage{amssymb}   % Common math symbols
\usepackage{makecell}
\usepackage{caption}
\usepackage{xcolor}

\begin{document}

%\title{Equilibrium with coordinate dependent diffusion: Comparison of different stochastic processes}
%\title{Symmetry and entropy of inhomogeneous diffusion: Understanding thermal equilibrium as It\^o-process}
\title{Equilibrium with coordinate dependent diffusion: Comparison of different stochastic processes }

\author{A. Bhattacharyay}
\email{a.bhattacharyay@iiserpune.ac.in}

\affiliation {\it Indian Institute of Science Education and Research, Pune, India}

\date{\today}

\begin{abstract}  
We show that, simultaneous local scaling of coordinate and time keeping the velocity unaltered is a symmetry of an It\^o-process. Using this symmetry, any It\^o-process can be mapped to a universal additive Gaussian-noise form. We use this mapping to separate the canonical and micro-canonical part of stochastic dynamics of a Brownian particle undergoing coordinate dependent diffusion. We identify the equilibrium distribution of the system and associated entropy induced by coordinate dependence of diffusion. Equilibrium physics of such a Brownian particle in a heat-bath of constant temperature is that of an It\^o-process.

\end{abstract}


\maketitle 

Thermodynamic equilibrium of a confined Brownian particle (BP) with coordinate-dependent diffusion needs to be understood for the possibility of its extensive applications \cite{roussel2004reaction,barik2005quantum,sargsyan2007coordinate,chahine2007configuration,best2010coordinate,lai2014exploring,berezhkovskii2017communication,foster2018probing,ghysels2017position, yamilov2014position}. A BP near a wall or interface undergoes coordinate-dependent diffusion as a function of its distance from the interface. Diffusivity of a BP gradually reduces from its bulk value, due to hydrodynamic effects, as it approaches an interface \cite{faucheux1994confined,Faxen1924,Brenner1961TheSM,Cox1967TheSM}. Models for such systems involve stochastic differential equation (SDE) with multiplicative noise. Solving these models requires adoption of a particular convention for stochastic integration as suggested by It\^o, Stratonovich or, for that matter, many other variants of Stratonovich's convention. 

It was Kiyosi It\^o who (in 1941) developed the method (convention) to solve stochastic differential equations with multiplicative-noise. In It\^o's convention, there does not arise any temporal correlation in the stochastic term and, it preserves thermal nature of the noise term \cite{ito1942differential,ito1986kiyosi}. About twenty-five years later, Stratonovich proposed a different convention \cite{strat} to get another solution to the same problem. Following this, It\^o has done some extensive work on Stratonovich's convention and has shown a correspondence to exist between these conventions.   

Consider a standard over-damped Langevin dynamics for a diffusion process of a BP in one-dimension:
\begin{equation}
\frac{dx}{dt}=-\frac{1}{\Gamma(x)}\frac{\partial V(x)}{\partial x}+\sqrt{\frac{2k_BT}{\Gamma(x)}}\eta(t).
\end{equation}
Here, $V(x)$ is the confining potential to the BP at position $x$. The function $\Gamma(x)$ is coordinate-dependent damping coefficient related to the local diffusivity by fluctuation-dissipation relation (FDR) $D(x)=\frac{k_BT}{\Gamma(x)}$ where $k_B$ is the Boltzmann constant. This relation defines the constant temperature $T$ of the system. Presence of the FDR ensures relaxation of the BP of mass $m$ to local equilibrium within a relaxation time scale $m/\Gamma(x)$ at position $x$. In equation (1), $\eta(t)$ is a standard Gaussian white noise of unit strength. Corresponding Fokker-Planck equation to this Langevin dynamics would admit an equilibrium distribution of the form $\rho(x) = Ne^{-V(x)/k_BT}$ (where $N$ is a constant) neither by It\^o's convention nor by that of Stratonovich's. 

Which convention (It\^o's or Stratonovich's) does represent equilibrium statistics of a {\it given Langevin dynamics} remains an open issue till date. Langevin dynamics is the generalization of Hamilton's equations to the stochastic domain and how to write it to include a heat-bath at constant temperature is well settled problem. N. G. van Kampen had elaborated, in the ref.\cite{van1981ito}, that one needs to understand different conventions (e.g., It\^o's or Stratonovich's) to represent different physical processes of space-dependent diffusion. However, understanding the physics of the It\^o-process, in this context, probably has suffered from lack of interest on the part of physicists who consider the Stratonovich's convention to be more physically motivated. 

There exist some arguments given in support of It\^o's convention \cite{bhattacharyay2019equilibrium,bhattacharyay2020generalization,maniar2021random, bhattacharyay2022physical} representing the thermal equilibrium of such systems. In the existing literature, however, dominant arguments are in favour of {\it a priori} consideration of $\rho(x) = Ne^{-V(x)/k_BT}$ to be the equilibrium distribution which It\^o's convention does not result in. One uses almost any generalization of Stratonovich's convention to mathematically justify this {\it a priori} fixed distribution \cite{sancho1982adiabatic,lau2007state, sokolov2010ito, tupper2012paradox,sancho2011brownian, farago2014fluctuation, farago2014langevin, leibovich2019infinite} and, one does that by considering some variant of Stratonovich's convention being more appropriate for a physical process. The It\^o-vs-Stratonovich controversy is quite alive and is a matter of intense debate \cite{volpe2016effective, bo2019functionals, hottovy2012noise, yang2013brownian, bo2019functionals, mannella2022ito}

In this paper we establish the physical basis of It\^o's convention capturing the thermal process (which we call It\^o-process) that represents equilibrium of a BP with coordinate-dependent diffusion in a heat-bath of constant temperature. We show that all It\^o-processes can be mapped to an additive Gaussian white-noise form. This mapping separates the canonical equilibrium of an It\^o-process from its micro-canonical part to help clearly identify the density of states arising from coordinate-dependent diffusion. It is also shown that the mapping to additive noise form is part of a general symmetry of all It\^o-processes. 
	
The Fokker-Planck equation, corresponding to the Langevin dynamics (1), as obtained by It\^o's convention would result in the equilibrium distribution of position $\rho_I(x) = M\frac{D_0}{D(x)}e^{-V(x)/k_BT}$ where $M$ is a normalization constant. This equilibrium distribution written in terms of {\it Helmholtz free energy} of the system would include the {\it inhomogeneity-entropy} $S_I=k_B\log{D_0/D(x)}$, where $D_0$ denotes diffusion constant in bulk and $D_0/D(x)$ is the (dimensionless) density of states which is a ratio of $1/D(x)$ to $1/D_0$. We show that, the factor of (micro-canonical) $D_0/D(x)$ to the canonical Boltzmann-Gibbs measure $e^{-V(x)/k_BT}$ is a general feature of the equilibrium distribution of such inhomogeneous diffusion processes.  

Langevin dynamics modelling under-damped (general) time-evolution of a BP of mass $m$ with coordinate-dependent diffusion in the confinement of potential $V(x)$ is
\begin{eqnarray}\nonumber
&&  \frac{dx}{dt} = v \\\nonumber
&& m\frac{dv}{dt} = -m\zeta(x)v -\frac{\partial V(x)}{\partial x}+ \sqrt{2k_BTm\zeta(x)}\times\eta(t).\\
\end{eqnarray}
In this dynamics, $\Gamma(x)=m\zeta(x)$ is the space-dependent damping coefficient where $\zeta(x)^{-1}$ is the local relaxation time-scale. A BP being disturbed at time $t=0$ (say) comes to equilibrium, on average, within the relaxation time. Over-damped limit defined by $m\to 0$ and $\zeta(x)\to\infty$ keeping $\Gamma(x)$ finite being taken on eqn.(2) results in eqn.(1) in the absence of inertia term.

Consider scaling of space and time by the function $f[x(t)]\equiv f(x)$ at time $t$, such that $dx^\prime \to f(x)dx = d\phi(x)$ and $dt^\prime \to f(x)dt$. Also consider that $\phi(x) = \int{f(x)dx} =x^\prime$ being defined, $\phi^{-1}(x^\prime)=x$ exists. For example, when $f(x)=\zeta(x)$ where $\zeta(x)$ is single valued, $x^\prime=\phi(x)=\int{\zeta(x)dx}$ remains defined. In this method of scaling, the velocity remains unaltered which is crucial in the sense that the thermal scale of energy (or energy in general) does not change under the scaling. Scaling of time by $\zeta(x)^{-1}$ is choosing the unit of time to be the local relaxation time under diffusion. In what follows, we will make use of this particular scaling function $f(x)=\zeta(x)$ to reveal the physics of It\^o-process. 

By scaling coordinate and time with a well defined function $f(x)$, one can rewrite Langevin dynamics (2) as
\begin{eqnarray}\nonumber
&& \frac{dx^\prime}{dt^\prime} = v \\\nonumber
&& m\frac{dv}{dt^\prime} = -m\frac{\zeta^\prime(x^\prime)}{f^\prime(x^\prime)}v - \frac{\partial V^\prime(x^\prime)}{\partial x^\prime}+ \sqrt{2k_BTm\frac{\zeta^\prime(x^\prime)}{f^\prime(x^\prime)}}\eta(t^\prime)\\
\end{eqnarray}
where a factor of $\sqrt{1/f(x)}$ has been absorbed in $\eta(t^\prime)$. This transformation of $\eta(t)\to\eta(t^\prime)$ must be done keeping in mind that $\eta(t)$ scales as $1/\sqrt{dt}$.  In eqn.(3), $f^\prime(x^\prime)\equiv f[\phi^{-1}(x^\prime)]$, $\zeta^\prime(x^\prime)\equiv \zeta[\phi^{-1}(x^\prime)]$ and $V^\prime(x^\prime)\equiv V[\phi^{-1}(x^\prime)]$ are the factions $f(x)$, $\zeta(x)$ and $V(x)$ written in new coordinate $x^\prime$. At this stage, these transformations do not alter the form of Langevin-dynamics needed in keeping with the FDR. In what follows, we will also see that an It\^o-process remains unaltered under these transformations.

With the specific choice of $f(x)=\zeta(x)$, one measures time in the natural time scale of diffusive transport i.e., $\zeta(x)^{-1}$ which helps map the given Langevin-dynamics (eqn.(2)) to a universal additive Gaussian-noise form,
\begin{eqnarray}\nonumber
&& \frac{dx^\prime}{dt^\prime} = v \\\nonumber
&& m\frac{dv}{dt^\prime} = -mv -\frac{\partial V^\prime(x^\prime)}{\partial x^\prime} + \sqrt{2k_BTm}\times\eta(t^\prime).\\
\end{eqnarray}
By choice of the scale $f(x)=\zeta(x)$, this mapping is unique to a given system characterised by $\zeta(x)$. The universal canonical form of homogeneous diffusive transport (eqn.(4)) is characterised by the constant temperature of the bath $T$ and constant relaxation time unity. 

The scaling of space has hidden, at this stage, the inhomogeneity of space within a micro-canonical density of states which will show up on returning to original scales. It will be explained in this letter that, such a mapping exists because the position uncertainty of inhomogeneous Brownian motion is entirely determined by the uncertainty in velocity of the BP which is fixed by the temperature of the system. 

The dynamics with additive noise (eqn.(4)) will equilibrate at a minimum of confining potential with the probability density
\begin{equation}
\rho[x^\prime,v] = N e^{-\frac{V^\prime(x^\prime)}{k_BT}} e^{-\frac{mv^2}{2k_BT}},
\end{equation}
at times $(t^\prime>>1)$, where $N$ is a normalization constant. This is the Boltzmann-Gibbs distribution of the BP in canonical equilibrium with the heat-bath at a constant temperature $T$.

Restoring the coordinates to normal scales by using $dx^\prime = \zeta(x)dx$, one can now rewrite the distribution as
\begin{equation}
\rho(x,v) = M\frac{D_0}{D(x)}e^{-\frac{V(x)}{k_BT}} e^{-\frac{mv^2}{2k_BT}},
\end{equation}
where $D(x)=k_BT/m\zeta(x)$, $D_0$ is the diffusion constant in bulk (away from interfaces). In the expression of probability density (eqn.(6)), $M$ is an overall normalization constant.

The distribution (eqn.(6)) includes the factor of density of states $\Omega(x)=D_0/D(x)$ to the Boltzmann-Gibbs measure in thermal equilibrium where $D_0>D(x)$. This same distribution (6) would result from corresponding SDE to eqn.(2) when treated by It\^o's convention. Let us call the distribution (6) to be {\it It\^o-distribution} after Kiyosi It\^o who had discovered the corresponding stochastic process in 1941. 

The distribution (6) can conveniently be rewritten in another form using Helmholtz free energy as
\begin{equation}
\rho(x,v) = M^\prime e^{-\frac{1}{k_BT}\left (H(x,p)-T[k_B\log{\frac{D_0}{D(x)}]}\right )},
\end{equation}
where $H(x,p)$ is the Hamiltonian of the system with momentum $p=mv$. This identifies $S[D(x)]=k_B\log{\frac{D_0}{D(x)}}$ as the {\it inhomogeneity entropy} corresponding to microcanonical degeneracy $\Omega(x)=D_0/D(x)$ of states at position $x$ in unit interval. In other words, $D_0/D(x)$ is the density of states originating from coordinate-dependent diffusion which breaks homogeneity of space even in the absence of any other field.

Let us have a look at the stochastic integral equation corresponding to the dynamics of velocity in eqn.(2). We pay attention to the velocity part alone because that is what matters. In the integral form of the SDE, the noise term is well defined as a continuous Brownian motion without pathologies of a white noise. The equation is
\begin{eqnarray}\nonumber
&& m(v(t)-v(0)) = \int_0^t{-mv\zeta(x)ds} -\int_0^t{\frac{\partial V(x)}{\partial x}} ds\\ && + \int_0^t{\sqrt{2k_BTm}\times\sqrt{\zeta[x(s)]}\times dW(s)},
\end{eqnarray}
where $dW(s)=\eta(s)ds$ is increment in the process of Brownian motion. This $dW(s)$ scales as $\sqrt{(ds)}$ as is obvious for any diffusion process. The change of time scale $ds^\prime \to \zeta[x(s)]ds$ in the Langevin dynamics would correspond to
\begin{eqnarray}\nonumber
dW^\prime(s) &&=\sqrt{\zeta[x(s)]}dW(s)\\ &&=Lt_{\delta s\to 0}\sqrt{\zeta[x(s)]}[W(s+\delta s)-W(s))],
\end{eqnarray}
in the stochastic integral (last term in (8)). According to It\^o-convention, the integral is sum of infinitesimal increments $\sqrt{\zeta[x(s)]}[W(s+\delta s)-W(s))]$ from $0$ to $t$ \cite{oksendal2013stochastic,gardiner1985handbook}. Thus, the scaling is consistent with an It\^o-process as $\sqrt{\zeta[x(s)]}$ enters each increment with its value at start of the interval $\delta s$ at $s$.

In eqn.(3), form of the stochastic term in relation to damping keeps FDR manifestly present. Such a scaling of time, termed as {\it random time change}, is detailed in ch. 8.5 in \cite{oksendal2013stochastic} by \O{}ksendal. Under this scaling, the damping $\zeta(x)$ locally transforms to $\zeta(x)/f(x)$. On solving eqn.(3) by It\^o's convention followed by transformation back to the normal scales would have factors $1/f(x)$ cancel out $f(x)$ in the density of states to result in the same distribution (6). Therefore, the local transformation of coordinate and time by any function $f(x)$ keeps an It\^o-process invariant. This transformation is a symmetry of It\^o-process. On the contrary, Stratonovich or generalized-Stratonovich processes would not possess this particular symmetry.

Estimation of minimum volume for a BP in velocity-position phase-space would explain the density of states $D_0/D(x)$. Consider homogeneous (bulk) Brownian motion for the sake of simplicity. Brownian motion is a process which has an average minimum time scale - the relaxation time scale $1/\zeta$ - below which the motion is not defined as Brownian. Spread of the velocity distribution of the BP at temperature $T$ is $\Delta v = \sqrt{k_BT/m}$ for a BP of mass $m$. Uncertainty in position of the BP in the relaxation-time interval is, therefore, $\Delta x=\Delta v/\zeta$. This gives the phase space volume corresponding to the minimum time interval $1/\zeta$ to be $\Delta v\Delta x = k_BT/m\zeta = k_BT/\Gamma = D_0$. This is the minimum phase volume available to the BP in its phase space and this phase-volume stands for a single state. One may recall, in quantum mechanics, this minimum phase volume is Planck's constant $h$ in position-momentum space.

Exactly by the same logic based on local relaxation time $1/\zeta(x)$ for coordinate-dependent diffusion, local diffusivity $D(x)$ is the minimum phase space volume available to a BP at position $x$. Therefore, at $x$, there are $D_0/D(x)$ states available to a BP within phase-volume $D_0$ as opposed to only one state in bulk to the $D_0$. This is the physics underlying larger density of states to a BP near an interface where $D(x)<D_0$. A BP would spend more time near an interface than in the bulk is an experimentally known fact \cite{carbajal2007asymmetry} which could be accounted for by the density of states in the theory of It\^o-process. For example, experimental results (insets of Fig.4 in the
ref.\cite{carbajal2007asymmetry}) given by Carbajal-Tinoco et al., show a clear shift of the peak of position distribution of a BP towards the confining walls where a relatively long tail of the distribution remains present towards the bulk of the fluid. Ref.\cite{carbajal2007asymmetry} explicitly mentions the fact that similar trend in position distribution is seen near the top as well as the bottom surface indicating the fact that, it is not an effect due to gravity. 

The {\it inhomogeneity-entropy},
\begin{equation}
S[D(x)] = k_B\log{\frac{D_0}{D(x)}},
\end{equation}
originating from coordinate-dependence of diffusion is positive definite because the bulk diffusion constant $D_0\geq D(x)$. A reasonable assumption that, with $T\to 0$ one must have $D(x)\to D_0$ as quickly as $D_0\to 0$ makes the {\it inhomogeneity-entropy} vanish at $T=0$ to keep it defined at all temperatures.

Langevin dynamics of BP in the over-damped limit (in the absence of correlations in noise) is given by the eqn.(1). Fokker-Planck equation to eqn.(1), according to It\^o's convention, is 
\begin{equation}
\frac{\partial \rho(x,t)}{\partial t} = -\frac{\partial}{\partial x}\left ( -\frac{\rho(x)D(x)}{k_BT}\frac{\partial V(x)}{\partial x}-\frac{\partial}{\partial x}\rho(x)D(x)\right).\\
\end{equation} 
In eqn.(11), the drift current density comes from $-\frac{1}{\Gamma(x)}\frac{\partial V(x)}{\partial x}$ in Langevin dynamics. The diffusion current density $-\frac{\partial}{\partial x}\rho(x)D(x)$ is universal i.e., it does not depend on what convention (e.g., It\^o or Stratonovich) is used. Had one used Stratonovich's convention, one would get exactly the same diffusion current density term along with some spurious drift current due to temporal correlation in noise.

On setting the probability current density to zero in accordance to detailed balance, this Fokker-Planck equation admits stationary equilibrium distribution for position 
\begin{equation}
\rho(x)=M\frac{D_0}{D(x)}e^{-\frac{V(x)}{k_BT}}.
\end{equation}
There exists example of arriving at a distribution function of the same structure as that in eqn.(12) in ref.\cite{ryter1981brownian, jayannavar1995macroscopic}. However, these works explicitly consider the pre-factor of $1/D(x)$ is arising due to a space-dependent temperature.

Dynamics given by eqn.(1) could also be brought to additive-noise form by local scaling of coordinate and time with $\Gamma(x)$. Easy to notice that, over-damped limit $m\to 0$ on the distribution (6) readily gives distribution (12). Taking the over-damped limit on eqn.(4) (i.e. $m\to 0$), where space and time have been scaled by $\zeta(x)$, is not possible because the other limit $\zeta(x)\to\infty$ could not be taken explicitly. However, the scaling and over-damped limit will commute if, instead, $\Gamma(x)$ (which remains finite) is chosen to be the scaling function. Choice of $\Gamma(x)$ as the scaling function is as good and equivalent but not used in the present context to explicitly scale time in its local unit set by diffusion.

In the presence of a coloured (anticipating) noise, the Fokker-Planck equation to eqn.(1), in general, has the structure
\begin{eqnarray}\nonumber
\frac{\partial \rho(x,t)}{\partial t} &=& -\frac{\partial}{\partial x}\left ( -\frac{\rho(x)D(x)}{k_BT}\frac{\partial V(x)}{\partial x}-\frac{\partial}{\partial x}\rho(x)D(x)\right)\\ &-&\alpha\frac{\partial}{\partial x}\rho(x)\frac{\partial}{\partial x}D(x),
\end{eqnarray}
where $\alpha$ is a constant. For a Stratonovich process $\alpha = 1/2$ and for a H\"anggi-Klimontovich (H-K) or anti-It\^o process, $\alpha = 1$. Corresponding to a Stratonovich process, the strength of coordinate dependent stochastic term (e.g., $\zeta[x(s)]$) in eqn.(9) between the interval $s$ to $s+\delta s$ is set to an average between the two of its values at $s$ and $s+\delta s$. Whereas, for a H-K process it is set to that at the time $s+\delta s$. This is what makes noise in these processes to be anticipating in nature and the noise is a coloured noise. Experimental results in ref.\cite{volpe2010influence,brettschneider2011force} are interpreted to have measured the spurious drift (as that in H-K process). The H-K process or anti-It\^o process might exist, however, it is a non-equilibrium process.

Stationary distribution for the H-K process $\rho(x) = Ne^{-V(x)/k_BT}$ is obtained by setting the probability current density to zero in eqn.(13). As another example, one may note that, similar Fokker-Planck equation is derived by Sancho et al., in 1982 (ref.\cite{sancho1982adiabatic}) using the method of adiabatic elimination. That the dynamics developed in ref.\cite{sancho1982adiabatic} is exactly the H-K (anti-It\^o) process could be understood by noting that eqn.(2.6) and (2.10) in ref.\cite{sancho1982adiabatic} employ Taylor expansions of damping backward in time. The damping and equivalently the noise-strength at an earlier time $t^\prime$ is obtained as an expansion over that at present time $t$ which makes it effectively involve the similar coloured noise.

Coloured noise, when used in over-damped dynamics, makes it an out of equilibrium process. In under-damped dynamics, coloured noise with a correlation-time (or microscopic collision time) $\tau_c$ is quite standard (e.g., chap.10 of the book by No\"elle Pottier \cite{pottier2009nonequilibrium} and chap.15 of the book by Federick Reif \cite{reif1965statistical}). This is standard because the second fluctuation dissipation theorem holds at the limit $\tau_c<<\zeta^{-1}$ and the noise remains thermal. 

However, the typical local time scale of dynamics in the over-damped regime is $\Gamma(x)$ which is finite where the relaxation time has been taken to satisfy the limit $\zeta(x)^{-1}\to 0$ everywhere in space. Therefore any temporal correlation of noise in the over-damped case is on a time scale much bigger than $\zeta(x)^{-1}$. This will break the second fluctuation dissipation relation in a hidden way one does not pay attention to that \cite{pottier2009nonequilibrium}. Therefore, an over-damped process with coloured noise is never an equilibrium process even when its stationary distribution is $\rho(x) = Ne^{-V(x)/k_BT}$. If one does understand this simple fact, then it would not be difficult to get the fallacy of using the It\^o-vs-Stratonovich controversy to promote $\rho(x) = Ne^{-V(x)/k_BT}$ as the equilibrium distribution under coordinate dependent diffusion. 

In this paper, we have identified the scaling of coordinate and time keeping velocity unaltered to be a symmetry of the It\^o-process. Due to this symmetry, to a given It\^o-process with multiplicative Gaussian white noise, there exists a unique mapping to homogeneous diffusion. Equipped with this we have identified the equilibrium distribution to an It\^o-process to be completely consistent with the Boltzmann-Gibbs measure up to a factor of a micro-canonical density of states. An {\it inhomogeneity-entropy} arising from coordinate dependence of diffusion gets identified. We prefer to call this {\it inhomogeneity entropy} because it is arising out of effective inhomoheneity of space and time. It is not due to symmetry and resulting degeneracy of the interactions. 

Diffusion becomes coordinate dependent in crowded environments like that inside a cell or in a porous medium. These systems are often near-equilibrium ones driven weakly to keep response of the system linearly related to the drive. For statistics under such conditions, all stationary moments are generally evaluated using equilibrium distribution of the system which has not got distorted much under weak driving. True thermal equilibrium corresponding to a constant maximum entropy is a rather restrictive situation which hardly gets maintained in real systems. The distribution obtained under strict conditions of thermal equilibrium finds its widest applications in all real systems in the linear response regime. Because of this generality to application, identifying the right equilibrium distribution of stochastic processes modelled by multiplicative Gaussian-noise is of enormous significance.

%\newpage
%\bibliography{reference.bib}

%

\end{document}